\documentclass{article}
\usepackage[T1]{fontenc}

\makeatletter

\providecommand{\LyX}{L\kern-.1667em\lower.25em\hbox{Y}\kern-.125emX\@}

\makeatother

\begin{document}

\title {The vortex oscillations 
in Abelian Higgs model}
\author {J. Karkowski 
	\thanks{Institute of Physics, Jagellonian University,
	30-064 Krak\'{o}w, Reymonta 4, Poland}
	\ and
	Z. \'{S}wierczy\'{n}ski
	\thanks{Institute of Physics, Pedagogical University, 
	Podchor\c a\.{z}ych 2, 30-084 Krak\'{o}w, Poland}}

\maketitle

\begin{abstract}
The excitations of the vortex 
in Abelian Higgs model with small ratio of vector and Higgs particle 
masses are considered. Three main modes encountered in numerical 
computations are described in detail. They are also compared to analytic 
results obtained recently by Arod\'{z} and Hadasz \cite{AH1}.
\end{abstract}

\section{Introduction}
\indent

The vortex type  solutions of nonlinear field equations 
are found to be important in many areas 
of physics. They are useful in considerations concerning 
different phenomena in field theory, cosmology and condensed matter 
physics \cite{GEN}. However both static and dynamic vortex configurations 
are known only approximately since they are usually described by 
complicated non-linear differential equations. Therefore the numerical 
methods must be applied to examine time-dependent vortex solutions.

The authors of paper \cite{AH1} have investigated an excitation of the vortex
in the Abelian-Higgs model with the mass of the Higgs field much bigger
than the mass of the vector particle.
They considered time dependent, axially symmetric fields with time and
radial components of the gauge potential equal to zero. In
the central part of the vortex the fields were approximated by polynomials. 
These
polynomials were continuously matched to the asymptotics describing
outer part of the vortex. The authors began with finding an 
approximation to the static vortex solution. Next they assumed that
the evolution of the Higgs field and the azimuthal component of the
gauge potential is frozen and considered harmonic oscillations 
of the gauge potential component parallel to the vortex axis. Finally
they found corrections to the Higgs field and the 
azimuthal component of the gauge potential.

In the present paper we would like to investigate described above excitation
of the vortex with the help of numerical computations. 
We have computed evolution
of axially symmetric fields with appropriatelly chosen initial data.
The approximation used in \cite{AH1} was simple,
and it led to the prediction of oscillations
of the Higgs field and of the azimuthal component of the gauge potential
with the frequency equal to the doubled frequency of the oscilations
of the gauge potential component  parallel to the vortex axis.
The solutions we have obtained are more complicated. Apart from the 
predicted mode, they contain also
oscillations corresponding to other vortex modes.

Our paper is organized as follows. In Sec. 2 we
introduce an axially symmetric ansatz and transform the equations to
the form which is more convenient in numerical computations.
Sec. 3 has rather technical character: containes description
of our computations. The results are presented and discussed in Sec. 4.
Finally Sec. 5 involves some general remarks and conclusions
summarizing our paper.

\section{Abelian Higgs Model}

The Abelian Higgs model is described by the following 
Euler-Lagrange equations

\begin{equation}
\label{start1}
(\partial _{\nu }+iqA_{\nu })(\partial ^{\nu }+iqA^{\nu })\Phi 
+\frac{\lambda }{2}\Phi (\mid \Phi \mid ^{2}-\frac{2m^{2}}{\lambda })=0 ,
\end{equation}

\begin{equation}
\label{start2}
\partial _{\mu }F^{\mu \nu }=iq\left( \Phi ^{*}\partial ^{\nu }\Phi 
-\Phi \partial ^{\nu }\Phi ^{*}\right) -2q^{2}A^{\nu }\mid \Phi \mid ^{2} .
\end{equation}

Our notation follows the paper \cite{AH1}. Here \( \Phi  \) is a complex
scalar field, \( A_{\nu } \) is \( U\left( 1\right)  \) gauge field, \( m \),
\( q \), and \( \lambda  \) are positive constants. The signature of the 
metric tensor is \( \left( +,-,-,-\right)  \) .

We restrict our considerations to the axially symmetric 
field configurations described by
the following Ansatz

\begin{equation}
\label{ansatz1}
\Phi =\sqrt{\frac{2m^{2}}{\lambda }}e^{i\theta }F\left( t,r\right) ,
\end{equation}

\begin{equation}
\label{ansatz2}
A_{0}=0,\, \, \, A_{3}=A\left( t,r\right) ,
\end{equation}

\begin{equation}
\label{ansatz3}
A_{1}=\frac{\sqrt{2}m}{q r}\sin\theta \left( 1-\chi \left( t,r\right) \right) ,
\end{equation}

\begin{equation}
\label{ansatz4}
A_{2}=-\frac{\sqrt{2}m}{q r}\cos\theta \left( 1-\chi \left( t,r\right) \right) ,
\end{equation}
where \( r=\sqrt{2m^{2}\left( (x^{1})^{2}+(x^{2})^{2}\right) } \),
\( \theta =\arctan \left( x^{2}/x^{1}\right)  \),
\( t=\sqrt{2}mx^{0} \). The fields have to be non - singular on the \( x^{3} \)
axis. This requirement implies

\begin{equation}
\label{non-singular}
F\left( t,0)\right) =0,\, \, \, \chi \left( t,0\right) =1,
\, \, \, \frac{d}{dr}F\left( r=0,t\right) =0 .
\end{equation}
The axially symmetric ansatz (\ref{ansatz1})-(\ref{ansatz4}) applied to the
equations (\ref{start1})-(\ref{start2}) simplifies them to the form

\begin{equation}
\label{F_rownanie}
\ddot{F}=F^{\prime \prime }+\frac{1}{r}F^{\prime }-\left( \frac{1}{r^{2}}\chi ^{2}+\frac{q^{2}}{m^{2}}A^{2}\right) +\frac{1}{2}\left( F-F^{3}\right) ,
\end{equation}
\begin{equation}
\label{chi_rownanie}
\ddot{\chi }=\chi \prime \prime -\frac{1}{r}\chi \prime -\kappa ^{2}F^{2}\chi ,
\end{equation}

\begin{equation}
\label{A_rownanie}
\ddot{A}=A^{\prime \prime }+\frac{1}{r}A^{\prime }-\kappa ^{2}F^{2}A ,
\end{equation}
where dot denotes the time derivative and prime the derivative with respect
to the variable r. The dimensionless parameter \( \kappa =\sqrt{2q^{2}/\lambda } \)
is equal to the ratio of the vector particle mass and the mass of the Higgs
particle. In this paper we restrict ourselves to small values 
of \( \kappa \, \left( 0.05,\, 0.1,\, 0.2\right)  \).
We also assume that \( q/m=1 \). This does not limit the generality of
our considerations since it can be achieved by rescaling the field \( A \).

Let us note that all field configurations 
of the form (\ref{ansatz1})-(\ref{ansatz4}) with finite energy 
per unit of length in the \( x^{3} \) direction have unit topological charge.
If in addition we assume that the functions \( f \) and \( \chi  \) 
are time independent
solutions of the equations (\ref{F_rownanie})-(\ref{chi_rownanie}) 
with \( A_{3}\equiv A\equiv 0 \)
then we obtain  the well known Nielsen-Olesen vortex. 
This static solution was examined
in many papers. However its exact analytical form is not known.
Only approximate methods (both analytical and numerical) have been worked out
and can be adopted as the starting point to the precise numerical calculations.
These are neccessary as the time dependent analysis is very subtle.

The equations (\ref{F_rownanie})-(\ref{A_rownanie}) are singular in \( r=0 \).
Therefore we apply to them the following transformation which removes the
singularity

\begin{equation}
\label{x_trans}
x=\frac{r^{2}}{1+r} ,
\end{equation}

\begin{equation}
\label{f_trans}
f(t,x)=\frac{F(t,r)}{r} ,
\end{equation}

\begin{equation}
\label{h_trans}
h(t,x)=\frac{\chi (t,r)-1}{r^{2}} ,
\end{equation}

\begin{equation}
\label{a_trans}
a(t,x)=A(t,r) .
\end{equation}
 Thus we obtain the equations of the form

\begin{eqnarray}
\label{f_eq}
\ddot{f} = \left( \frac{\left( 2+r\right) r}
{\left( 1+r\right) ^{2}}\right) ^{2}f^{\prime \prime }+
\left( \frac{2}{\left( 1+r\right) ^{3}}+
\frac{2+r}{\left( 1+r\right) ^{2}}\right) f^{\prime } \nonumber \\
  -\left( 2h+r^{2}h^{2}+a^{2}\right) f +
\frac{1}{2} f \left( 1 - r^{2} f^{2} \right) ,
\end{eqnarray}

\begin{equation}
\label{h_eq}
\ddot{h}=\left( \frac{\left( 2+r\right) r}{\left( 1+r\right) ^{2}}\right) ^{2}h^{\prime \prime }+\left( \frac{2}{\left( 1+r\right) ^{3}}+\frac{2+r}{\left( 1+r\right) ^{2}}\right) h^{\prime }-\kappa ^{2}f^{2}\left( 1+r^{2}h \right) ,
\end{equation}

\begin{equation}
\label{a_eq}
\ddot{a}=\left( \frac{\left( 2+r\right) r}{\left( 1+r\right) ^{2}}\right) ^{2}a^{\prime \prime }+\left( 2\frac{\left( 2+r\right) ^{2}}{\left( 1+r\right) ^{3}}-\frac{4+r}{\left( 1+r\right) ^{2}}\right) a^{\prime }-\kappa ^{2}r^{2}f^{2}a .
\end{equation}
Here prime denotes the derivative with respect to \( x \), \( r=\left( x+\sqrt{x\left( 4+x\right) }\right) /2 \).
Some disadvantage of the transformation (\ref{x_trans}) is that the neighbourhood
of the point \( r=0 \) is mapped into a very small \( x \) interval (\( dx/dr\mid _{r=0}=0 \)).
Therefore the detailed solution in this region has to be examined separately.

\section{The method
}

In this paper we investigate the oscillations of the vortex caused by the excitations
of the field \( A \). The first step is to have the precise solutions for the
functions \( f \) and \( \chi  \) in the static vortex configuration.  
In \cite{AH1} these fields in the central part of the vortex were approximated
by polynomials. These polynomials were continuously matched to the appropriate
asymptotics describing external part of the vortex. There also exist more precise
numerical approximations of the static vortex \cite{KS1} but their accuracy is not
sufficient too. Therefore we have started with one 
of these approximate methods and then 
in order to improve the solutions we have applied the average method, that is
we have examined the time evolution of the vortex alone averaging the functions
\( f \) and \( \chi  \) around their equilibrium state.

The next step is to excite the vortex. We have done this by adding a non-zero 
field \( A \).
Following \cite{AH1} we have used a configuration which gives harmonic oscillations 
of this field while the vortex fields are frozen. Substituting 
\(A(t,r) = \hat{a}(r) \cos \omega t \)
into (\ref{A_rownanie}) one obtains

\begin{equation}
\label{alfa_eq}
\hat{a}^{\prime \prime }+\frac{1}{r}\hat{a}^{\prime }-\kappa ^{2}F^{2}\hat{a} +\omega ^{2}\hat{a} =0 .
\end{equation}

As was noticed in \cite{AH1} this equation has the form of the one-dimensional 
Schr\"odinger equation and possesses
 at least one bound state. Rewriting it in the
 variable \( x \) 
we obtain 
\begin{equation}
\label{a_bound_eq}
\left( \frac{\left( 2+r\right) r}{\left( 1+r\right) ^{2}}\right) ^{2}\frac{d^{2}\hat{a}}{dx^2}+\left( 2\frac{\left( 2+r\right) ^{2}}{\left( 1+r\right) ^{3}}-\frac{4+r}{\left( 1+r\right) ^{2}}\right) \frac{d\hat{a}}{dx}-\left( \kappa ^{2}r^{2}f^{2}-\omega ^{2}\right) \hat{a}=0 .
\end{equation}
 On the uniform grid with the spacing \( \Delta  \) eq. (\ref{a_bound_eq})
leads to the following finite difference equations

\begin{equation}
\label{eq_difference1}
\frac{4}{\Delta }(\hat{a}_{1}-\hat{a}_{0})+\omega ^{2}\hat{a}_{0}=0 ,
\end{equation}

\begin{eqnarray}
\left( \frac{\left( 2+r\right) r}{\left( 1+r\right) ^{2}}\right) ^{2}\left( \frac{1}{\Delta ^{2}}\right) (\hat{a}_{k+1}-2\hat{a}_{k}+\hat{a}_{k-1}) \nonumber \\
+\left( 2\frac{\left( 2+r\right) ^{2}}{\left( 1+r\right) ^{3}}-\frac{4+r}{\left( 1+r\right) ^{2}}\right) \frac{1}{\Delta }(\hat{a}_{k+1}-\hat{a}_{k}) \nonumber \\ 
-\left( \kappa ^{2}r^{2}f^{2}-\omega ^{2}\right) \hat{a}_{k}=0, \ k=1,2,... 
\label{eq_difference2}
\end{eqnarray}
For small values of the parameter \( \kappa  \) the frequency \( \omega \) 
is approximately equal to \( \kappa  \) \cite{AH1}. Therefore we can 
solve the above difference equations 
(\ref{eq_difference1}), (\ref{eq_difference2})
where \( \omega  \) is replaced by \( \kappa . \) We choose arbitrary \( \hat{a}_{0} \)
and succesively determine \( \hat{a}_{1} \), \( \hat{a}_{2} \), ... . The graphes
of the function \( A(r)=a(r^{2}/(1+r)) \) obtained from formulae 
(\ref{eq_difference1}), (\ref{eq_difference2})
with \( \Delta =0.01 \) and the approximation of the function \( A \) obtained
in the paper \cite{AH1} for \( \kappa =0.1 \) are ploted in Fig 1. 

Now we are going to investigate the evolution of the excited vortex using numerical
computations. We solve the following finite difference equations 
\begin{eqnarray}
f_{j+1,k} & = & 2f_{j,k}-f_{j-1,k}+\left( \frac{3}{4}\right) ^{2}\left( \frac{(2+r)r}{(1+r)^{2}}\right) ^{2}(f_{j,k+1}-2f_{j,k}+f_{j,k-1})\nonumber \\
 &  & +\left( \frac{3}{4}\right) ^{2}\Delta \left( 2\frac{\left( 2+r\right) ^{2}}{(1+r)^{3}}+\frac{r}{\left( 1+r\right) ^{2}}\right) (f_{j,k+1}-f_{j,k})\label{ftime} \\
 &  & +\left( \frac{3}{4}\right) ^{2}\Delta ^{2}\left( \frac{1}{2}-2h_{j,k}-\frac{1}{2}r^{2}f_{j,k}^{2}-r^{2}h_{j,k}^{2}-a_{j,k}^{2}\right) f_{j,k}\nonumber ,
\end{eqnarray}
\begin{eqnarray}
h_{j+1,k} & = & 2h_{j,k}-h_{j-1,k}+\left( \frac{3}{4}\right) ^{2}\left( \frac{\left( 2+r\right) r}{\left( 1+r\right) ^{2}}\right) ^{2}(h_{j,k+1}-2h_{j,k}+h_{j,k-1})\nonumber \\
 &  & +\left( \frac{3}{4}\right) ^{2}\Delta \left( 2\frac{\left( 2+r\right) ^{2}}{\left( 1+r\right) ^{3}}+\frac{r}{\left( 1+r\right) ^{2}}\right) (h_{j,k+1}-h_{j,k})\nonumber \\
 &  & -\left( \frac{3}{4}\right) ^{2}\Delta ^{2}\kappa ^{2}(1+r^{2}h_{j,k})f_{j,k}^{2}\label{htime} ,
\end{eqnarray}
\begin{eqnarray}
a_{j+1,k} & = & 2a_{j,k}-a_{j-1,k}+\left( \frac{3}{4}\right) ^{2}\left( \frac{\left( 2+r\right) r}{\left( 1+r\right) ^{2}}\right) ^{2}(a_{j,k+1}-2a_{j,k}+a_{j,k-1})\nonumber \\
 &  & +\left( \frac{3}{4}\right) ^{2}\Delta \left( 2\frac{\left( 2+r\right) ^{2}}{\left( 1+r\right) ^{3}}-\frac{4+r}{\left( 1+r\right) ^{2}}\right) (a_{j,k+1}-a_{j,k})\nonumber \\
 &  & -\left( \frac{3}{4}\right) ^{2}\Delta ^{2}\kappa ^{2}r^{2}a_{j,k}f_{j,k}^{2}\label{atime} ,
\end{eqnarray}
 where \( \Delta  \) is the space between the grid points in the x-direction;
the space between time slices is \( 3\Delta /4 \). As the initial values for
the fields \( f \) and \( h \) at \( t=0 \) we take found earlier functions
corresponding to the static vortex solution. The initial values of the field
\( a \) are computed from the formulae (\ref{eq_difference1}), 
(\ref{eq_difference2}) with \( \hat{a}_{0}=0.1 \).
We also assume that initially the derivatives with respect to time of
all fields are equal to zero. 
In order to remove disturbances of the solution caused by the right boundary
of our grid we have performed computations on the larger area and then we removed
the part of the grid which could be disturbed. 

\section{Numerical results}

We have computed the evolution of the vortex for \( 0\leq t\leq 600 \) and
\( \kappa =0.05,0.1,0.2 \) taking \( \Delta =0.01 \). The field \( A \) oscillates
with the frequency \( \omega \approx \kappa  \). Fig.2-Fig.7  present 
the evolution
of the functions \( \delta f_{x}(t)=f(t,x)-f(0,x) \), \( \delta h_{x}(t)=h(t,x)-h(0,x) \)
(i.e. the differences of the time-dependent and static values of the fields
\( f \) and \( h \) respectively) at the points \( x=5 \) and \( x=160 \).
We have chosen these values of the variable \( x \) to compare the oscillations
near the vortex core with those far from it.
The authors of the
paper \cite{AH1} predict oscillations of the field \( f \) and \( \chi  \) with
the frequency \( \omega =2\kappa  \) but they have only analysed the modes
forced by the oscillations of the field  \( A \). The numerical results we have
obtained are more complicated and involve also other vortex modes. These modes
will be presented below in more detail. It is also interesting that while the
oscillations of the Higgs field \( f \) start at once for all values of \( x \)
the oscillations of the field \( h \) are delayed for points far from the vortex
centre. They begin at the vortex core and then  propagate outside.
This fact is also illustrated in Fig.8-Fig.10 which show the plots of the functions
\( \delta h_{x} \) for fixed time values \( t=100,200,300 \). One can see
that in the region \( r>t \) the function \( \delta h_{x} \) is almost equal
to zero. It starts to oscillate only when the disturbance of the field \( h \)
arrives from the vortex centre. 

Let us now present the frequencies of the vortex oscillations in the neighbourhood
of its equilibrium state. We have computed the fourier transformations for the
functions \( \delta f_{x} \) and \( \delta h_{x} \) for three different values
of \( \kappa  \) mentioned above eg. \( \kappa =0.05,0.1,0.2 \). These calculations
have been performed
 for two distances from the vortex core. As before we have choosen points \( x=5 \) and \( 160 \).
We have presented our results for the
Fourier transformations in Fig.11-Fig.16.
 The most important  conclusions can be summarized
as follows. The main frequency of the vortex oscillation is equal to 
\( 2\kappa  \). The
appropriate peak can be easily observed in each of the figures Fig.11-Fig.16.
 These oscillations are
explained in detail in \cite{AH1} on the basis of the approximate analytic calculations.
They are forced by the corresponding oscillations of the field \( A \) with
frequency \( \kappa  \).
But the other vortex modes are also excited.
Let us stress that
their frequencies do not depend on the parameter \( \kappa  \). The approximate
numerical values of these frequencies are correspondingly equal to the mass of
the Higgs particle (\( 1.0 \) in our case) and \( 90\% \) of this mass (that
is \( 0.9 \)). Let us note that these values seem to be independent from
the field \( a \) and
are two points in the whole spectrum
of frequencies of the vortex oscillations computed in \cite{GH1}. However it is
interesting that the lower frequency was not detected for the field \( f \)
far from the vortex core (Fig.12).

\section{Ending Remarks}

The excitation of the vortex we have investigated was obtained by choosing
some particular initial data for the component \( A_z \) of the gauge field 
parallel to the vortex axis. In the static vortex solution this component
of the gauge field is equal to zero; in our case it oscillates with the
frequency approximately equal to \( \kappa \): the ratio 
of the vector boson and Higgs masses. 
The other fields also oscillate. Performing the Fourier 
transformation we found that the frequencies of these oscillations are grouped
near \(2 \kappa, 0.9, 1.0 \). The first value was predicted in \cite{AH1}
and is caused by the term proportional to \( A^{2} \) in (\ref{F_rownanie}).
The oscillations with the second frequency appear mainly on the vortex
core and are probably concerned with the bound state of the static
vortex solution \cite{GH1}. The third frequency simply equals to the mass
of the Higgs particle. However it should be noted that our computations
include about ten periods of the field \( A \) oscillations and therefore
the question concerning the long time behaviour of the system is still open.
The field configuration we have considered was constructed in such a way
that the field \( A \) starts its oscillations at once in the whole space.
These oscillations excite the Higgs field for all values of the spatial
coordinate. However the motion of the azimuthal component of the gauge field
looks differently. It begins at the vortex centre and then propagates
outside.

\vskip 1.cm

{\bf Acknowledgements.}\\
This work was supported in part by KBN grant No. 2 P03B 095 13.

\pagebreak

\end{document}


\noindent 
Fig. 1. The initial state of the field \( A \) - numerical and
	polynomial approximations.
\vskip .4cm
\noindent  	
Fig. 2. The difference \( \delta f_{x}(t) \) of the dynamical and static values 
	of the field \( f\) for \( x=5.0 \) and \( \kappa =0.1 \).
\vskip .4cm
\noindent
Fig. 3. The difference \( \delta f_{x}(t) \) of the dynamical and static values 
	of the field \( f\) for \( x=160.0 \) and \( \kappa =0.1 \).
\vskip .4cm 
\noindent
Fig. 4. The difference \( \delta h_{x}(t) \) of the dynamical and static values 
	of the field \( h\) for \( x=5.0 \) and \( \kappa =0.1 \).
\vskip .4cm 
\noindent
Fig. 5. The difference \( \delta h_{x}(t) \) of the dynamical and static values 
	of the field \( h\) for \( x=160.0 \) and \( \kappa =0.1 \).
	Note the delay time \( \approx x \). 
\vskip .4cm 
\noindent
Fig. 6. The difference \( \delta h_{x}(t) \) of the dynamical and static values 
	of the field \( h\) for \( x=160.0 \) and \( \kappa =0.05 \).
	Note the delay time \( \approx x \). 
\vskip .4cm 
\noindent
Fig. 7. The difference \( \delta h_{x}(t) \) of the dynamical and static values 
	of the field \( h\) for \( x=160.0 \) and \( \kappa =0.2 \).
	Note the delay time \( \approx x \). 
\vskip .4cm 
\noindent
Fig. 8. The snapshot of the difference \( \delta h_{x} \) for \( t=100.0 \).
\vskip .4cm 
\noindent
Fig. 9. The snapshot of the difference \( \delta h_{x} \) for \( t=200.0 \).
\vskip .4cm 
\noindent
Fig. 10. The snapshot of the difference \( \delta h_{x} \) for \( t=300.0 \).
\vskip .4cm 
\noindent
Fig. 11. The Fourier transformations modules 
	of the difference \( \delta f_{x}(t) \)
	for \( x=5.0 \) and \( \kappa=0.05, 0.1, 0.2\).
\vskip .4cm 
\noindent
Fig. 12. The Fourier transformations modules 
	of the difference \( \delta f_{x}(t) \)
	for \( x=160.0 \) and \( \kappa=0.05, 0.1, 0.2 \).
\vskip .4cm 
\noindent
Fig. 13. The Fourier transformations modules 
	of the difference \( \delta h_{x}(t) \)
	for \( x=5.0 \) and \( \kappa=0.1, 0.2 \).
\vskip .4cm 
\noindent
Fig. 14. The Fourier transformations modules 
	of the difference \( \delta h_{x}(t) \)
	for \( x=160.0 \) and \( \kappa=0.1, 0.2 \).
\vskip .4cm 
\noindent
Fig. 15. The Fourier transformation modules 
	of the difference \( \delta h_{x}(t) \)
	for \( x=5.0 \) and \( \kappa=0.05 \).
\vskip .4cm 
\noindent
Fig. 16. The Fourier transformation modules 
	of the difference \( \delta h_{x}(t) \)
	for \( x=160.0 \) and \( \kappa=0.05 \).